\documentstyle[art12a,12pt]{article}
\newcommand{\bqn}{\begin{eqnarray}}
\newcommand{\eqn}{\end{eqnarray}}
\newcommand{\nn}{\nonumber}

\newcommand{\LL}{{\cal L}}

\newcommand{\VV}{{\cal V}}

\newcommand{\gc}{\gamma^{5}}

\newcommand{\ra}{\rightarrow}
\newcommand{\ee}{{\rm e}}
\newcommand{\dis}{\displaystyle}
\newcommand{\bra}{\langle}
\newcommand{\ket}{\rangle}

\newcommand{\DEMU}{\partial^{\mu}}

\newcommand{\psib}{\bar{\psi}}

\newcommand{\dslash}{\hat{\partial}}

\newcommand{\fpi}{f_{\pi}}

\newcommand{\unmez}{{1\over 2}}

\newcommand{\intkk}{\int {d^2 k\over (2\pi)^2}}
\newcommand{\intkl}{\int\limits_{0}^{+\infty}dk}
\newcommand{\qd}{^{2}}
\newcommand{\re}{{\rm Re\;}}

\newcommand{\tr}{{\rm Tr\;}}

\newcommand{\fp}{\varphi_{\pi}}
\newcommand{\mo}{ m_0}
\newcommand{\ww}{\omega^2}

\newcommand{\kk}{k^2}
\newcommand{\eps}{\ee ^{(k+\eta)/r}}
\newcommand{\ems}{\ee ^{(k-\eta)/r}}
\renewcommand{\ss}{\sigma^2}
\renewcommand{\gg}{g^2}
\newcommand{\bs}{\hat{\sigma}}

\newcommand{\so}{\sigma_{0}}
\newcommand{\sso}{\sigma_{0}^2}

\newcommand{\jc}{{\cal J}_{\mu}^{5}}
\newcommand{\mpi}{m^2_{\pi}}

\newcommand{\hfb}{\hfill\break}

\def\tr{{\rm tr}}

\def \dalamb {\Box}
\begin{document}
\thispagestyle{empty}
\vspace*{4cm}
\begin{center}
  \begin{Large}
  \begin{bf}
   Thermodynamics \\
   of the massive Gross-Neveu model$^*$\\
  \end{bf}
  \end{Large}
  \vspace{15mm}
  \begin{large}
A. Barducci, R.Casalbuoni, M. Modugno and G. Pettini\\
  \end{large}
Dipartimento di Fisica, Univ. di Firenze\\
I.N.F.N., Sezione di Firenze\\
  \vspace{10mm}
  \begin{large}
R. Gatto\\
  \end{large}
D\'epartement de Physique Th\'eorique, Univ. de Gen\`eve\\
  \vspace{10mm}
\end{center}
  \vspace{2cm}
\begin{center}
UGVA-DPT 1994/06-854\\
\end{center}
\vspace{3cm}
\noindent
$^*$ Partially supported by the Swiss National Foundation

\newpage
\thispagestyle{empty}
\begin{quotation}
\vspace*{5cm}
\begin{center}
  \begin{Large}
  \begin{bf}
  ABSTRACT
  \end{bf}
  \end{Large}
\end{center}
  \vspace{5mm}
\noindent

We study the thermodynamics of massive Gross-Neveu models with explicitly
broken discrete or continuous chiral symmetries for finite temperature and
fermion densities. The large $N$ limit is discussed bearing attention to the
no-go theorems for symmetry breaking in two dimensions which apply to the
massless cases. The main purpose of the study is to serve as analytical
orientation for the more complex problem of chiral transition in
$4-$dimensional QCD with quarks. For any non-vanishing fermion mass
we find, at finite densities, lines of first order phase transitions. For
small mass values traces of would-be second order transitions and a
tricritical point are recognizable. We study the thermodynamics of these
models, and in the model with broken continuous chiral symmetry we
examine the properties of the pion like state.

\end{quotation}
\newpage

\sect{Introduction}

We have examined the massive Gross-Neveu model \cite{GN} at finite
temperature and density within the mean field approximation. In spite
of the problems related to the low dimensionality
of the model, this may represent in our opinion
a guide to the thermodynamics of chiral symmetry
restoration in QCD.
We have studied the model with a bare
mass term included, always kept non vanishing.
Due to such a choice, no chiral phase transition is
present, since chiral symmetry is
explicitly broken from the beginning and thus, strictly speaking,
the Mermin-Wagner-Coleman theorem \cite{MW}\cite{Co}
does not apply.
Although the symmetry is explicitly broken we find, as in our
previous study for a QCD model \cite{BC}, that some first order
phase transition still survives. The critical line obviously
moves in the plane of temperature and chemical potential
for growing masses, but it survives even for large mass values.
We derive the equation of state and study the phase
diagram for different choices of thermodynamical variables.
We discuss the isotherms in the pressure-inverse density plane,
which resemble the Van der Waals isotherms for the vapour-liquid
transition in the water. The construction is made through the
study of the effective potential, which contains
all the physical information
about the stable and metastable phases of a given model.
We can define a critical point as the ending point
of the coexistence region. There are, below this point,
two regions of very low
and of very high compressibility, separated by the coexistence region which
ends at the critical point. \hfb
We have also considered the explicit (small) breaking of
a continuous symmetry in order to implement the study of
soft-pion type properties
in the model. It is evident that (differently from $4-$dimensional QCD)
in two dimensions the pion-like particle cannot be
considered as a Goldstone particle
due to the Coleman theorem. Thus one principal difference between
previous studies \cite{BC} and the present work is that here
the zero mass limit cannot be taken without a
complete change of the picture.\hfb
To further comment on how no-go theorems work in a
finite temperature field theory and to better
motivate our presentation, we start by summarizing
in a very schematic way known results
in Section 2.\hfb
Section 3 is devoted to the results for the phase diagram and
the equation of state of the model with a broken discrete symmetry.
Here also naive zero mass limit results will be presented, to better
clarify those for the massive case in which we are indeed
interested, and at the
same time to show what would be reasonable to expect in analogous studies
in 3+1 dimensions.
In Section 4 we attempt at a description of pion properties in the model
with a broken continuous symmetry. As the effective potential in the mean field
approximation can be put
in the same form as that of  Section 3, the results
concerning the equation of state and the phase diagram are the same.
Finally, some useful calculations are summarized in the Appendix.\hfb

\sect{General review}

The Gross-Neveu model \cite{GN} is a well-known two-dimensional
theory with four-fermion interactions which is asymptotically
free. The fermion field has $N$ components. The model was
originally considered in the $1/N$ expansion.
The massless formulations, with  discrete
or continuous symmetry, have been extensively studied for zero or
for finite temperatures and densities \cite{Ja}-\cite{RW}, giving
rise to several discussions related to the low
dimensionality of the model, with its implications for
symmetry breaking and phase transitions, and to the validity
of the $1/N$ expansion (many aspects have been
already discussed in ref. \cite{RW}).\hfb

\noindent
\subsection{Discrete symmetry}

Let us summarize in a very schematic way the main
known results. The Lagrangian is
\bqn
{\cal L}={\bar \psi}i{\hat\partial}\psi
+{1\over 2}g^2\left({\bar\psi}\psi\right)^2
\label{gnlag}\eqn
(for studies of the limit $N\rightarrow\infty$ one will also define
$g^2 N=\lambda$). It is
invariant under the discrete chiral symmetry
\bqn
\psi\rightarrow \gamma_{5}\psi
\eqn
The four fermion interaction can be conveniently studied \cite{GN} by
introducing a $\sigma$ field in the generating functional
which satisfies the classical equation of motion $\sigma=g{\bar\psi}\psi$, and
transforming as $\sigma\rightarrow -\sigma$ under chiral symmetry.
The Lagrangian becomes
\bqn
{\cal L}= {\bar \psi}i{\hat\partial}\psi
-{1\over 2}\sigma^2 + g\sigma {\bar\psi}\psi
\eqn
and by integrating over the fermion field $\psi$
one can study the effective action as a functional of $\sigma$.\hfb

\noindent
a) $T=0,\mu=0$ \hfb

\noindent
At zero temperature $T$ and chemical potential $\mu$ the calculation indicate
the vacuum expectation value
$\langle\sigma\rangle\not= 0$, and therefore the discrete
chiral symmetry is spontaneously broken. This happens for any value of
the coupling constant.
There is no contradiction with the no-go theorems
\cite{MW}\cite{Co}.\hfb\vfill\eject

\noindent
b) $T\not= 0$, $\mu=0$ \hfb

\noindent
In this case the time dimension is bounded by $\beta=1/T$. The
thermodynamical limit can be taken only on the space dimension
(unless $\beta\rightarrow \infty$ which goes back
to case a)).
Thus, as far as the occurrence of spontaneous symmetry breaking is concerned,
the model behaves as one dimensional and the discrete chiral symmetry
is immediately restored at $T_{c}=0^{+}$. This is a manifestation
of the Mermin-Wagner theorem \cite{MW}.
The restoration of chiral symmetry at finite temperature
is driven by the presence of kinks and antikinks. These are
non constant field configurations connecting the two degenerate minima,
whose number grows with the volume \cite{DM}.
This happens for any large, finite $N$. If however
the limit $N\rightarrow\infty$ is taken before the
thermodynamical limit, these configurations are suppressed,  and one is
left within the mean field theory, where the model exhibits
a second order phase transition at a critical temperature
$T_{c}\not= 0$ \cite{Ja}.\hfb

\noindent
c) $T\not=0,\mu\not=0$\hfb

\noindent
By considering the model in the $N\rightarrow\infty$ limit, one can derive
analytically the phase diagram
in the plane of chemical potential and temperature $(\mu,T)$.
It turns out that there exists
a tricritical point separating  second order  phase transitions from
first order ones \cite{Wo}.
This result has been criticized by a lattice study \cite{KK} where the
authors find that at any $\mu\not= 0$ the phase transition is first order.
They claim that this is due to the formation
of kink-antikink configurations which are now not suppressed and trigger the
phase transition in this case. \hfb

\noindent
In our study of the discrete symmetry model at finite temperature and density
a bare mass is included, always taken different from zero. This ensures
the elimination of kink-antikink configurations, which are suppressed
in the thermodynamical limit (even in presence of a finite chemical
potential \footnote{One can convince oneself about this fact by looking
in ref. \cite{DM} at the procedure leading to the expression for the
number of kinks, which is independent of the chemical potential.}).\hfb
The model has been considered in the $N\rightarrow\infty$
limit and we believe that a $1/N$ expansion would not destroy the
qualitative picture provided the mass is always kept different from zero.\hfb

\noindent
\subsection{Continuous symmetry}

We summarize now the main results for the model with continuous
symmetry. One starts from the Lagrangian\vfill\eject
\bqn
{\cal L}={\bar \psi}i{\hat\partial}\psi
+{1\over 2}g^2\left({\bar\psi}\psi\right)^2
-{1\over 2}g^2\left({\bar\psi}\gamma_{5}\psi\right)^2
\eqn
which is invariant under the continuous chiral transformation
\bqn
\psi\rightarrow e^{i\theta\gamma_{5}}\psi
\label{transfcont}\eqn
In this case one introduces in the generating
functional, besides the scalar
field $\sigma$ also a pseudoscalar field
satisfying the classical equation of motion
$\pi=ig{\bar\psi}\gamma_{5}\psi$ \cite{GN}. The $\sigma$ and $\pi$
fields transform under the continuous chiral transformation as
\bqn
{\sigma\choose\pi}\ra\pmatrix{\cos 2\theta&\sin 2\theta\cr
-\sin 2\theta&\cos 2\theta\cr}{\sigma\choose\pi}
\label{transf}\eqn
Integrating over the fermion fields one gets the effective
action as a functional of the two fields $\sigma$ and $\pi$.\hfb

\noindent
a) $T=\mu=0$\hfb

\noindent
Due to chiral invariance, the effective potential is
a functional of $\rho^2\equiv \sigma^2+\pi^2$ only. Thus one
easily realizes that in the mean field approximation
its analytical expression is equivalent
to that of the discrete symmetry model. Then chiral symmetry appears to
be spontaneously broken. This is also the result at the leading order
in $1/N$ since there is no kinetic term for the collective fields.
Anyway, rigorously speaking, continuous chiral
symmetry cannot be broken since there are no Goldstone
bosons in two dimensions \cite{Co}. This shows up
at the next-to-leading order in $1/N$, due to IR
divergencies in the $\pi$ correlation function.
Nonetheless, as shown in refs. \cite{Wi}\cite{RW}, by choosing
a polar representation for the physical fields, longitudinal
and transverse
\bqn
\sigma+i\pi=\rho ~ e^{i\theta}
\label{polar}\eqn
one finds that the system still allows for an "almost long range order", since
the two-point correlation function
decreases with a power law for large distances,
the exponent being $1/N$ \cite{BK}\cite{Wi}.
Indeed this behaviour, which was originally found in the
low-temperature phase of the continuous symmetry X-Y spin model by Kosterlitz
and Thouless \cite{BK}, is simply dictated by dimensional
arguments. Actually it comes out from the fact that one is
considering the transverse fluctuations of the order parameter,
which in the IR are dominated by the free propagator of the
$\pi$ field, which is $\sim 1/k^2$ \cite{GN}\cite{RW}.
Thus the Fourier transform in two dimensions diverges
logarithmically for large distances (the same holds
for the $\theta$ propagator). Anyway, by using
the physical representation (\ref{polar}) this behaviour has
to be exponentiated.
As the hypotheses are general, the Kosterlitz-Thouless phase
appears simply as the "two dimensional version of long range order"
in a class of models \cite{DD}.\hfb\vfill\eject
A similar phenomenon occurs in 2+1 dimensions for massless bosonic
theories at finite temperature, as it appears from the
behaviour of the free two point
correlation function. (see Appendix D).\hfb

\noindent
b) $T\not=0$\hfb

\noindent
Again, for large but finite $N$, at $T\not= 0$, one can
easily find that the system behaves as in one dimension and that
even this "almost long range order" disappears (see Appendix D).
Anyway, even in this case, the $N\rightarrow\infty$ limit
can be formally taken in a way to eliminate any space-time
dependence of the correlation function. Thus one is left
within the mean field description, which
gives the same critical temperature as in the discrete symmetry model.
This solution seems to contradict the
Mermin-Wagner-Coleman theorems. We are going to comment on this
point in the following.\hfb

\noindent
It is evident that whenever non constant field configurations change
the picture provided by the mean field approach, the $1/N$ expansion
fails. In fact the fluctuations can be suppressed only if $N$ is taken
strictly infinite, rather than approximating with its leading term
a $1/N$ expansion.
Thus, if $N$ is large but finite, the final result is simply that
the symmetry can be broken spontaneously only if ${\cal D}>1$
(discrete symmetry) or ${\cal D}>2$ (continuous symmetry), ${\cal D}$
being the number of dimensions for which one takes the thermodynamical
limit.
It is clear that going from a zero temperature to a
finite temperature field theory, ${\cal D}$ decreases of one unity (this
will be further evidenced in the following and in Appendix D).\hfb
To specify something more about the
infinite $N$, let us recall some qualitative arguments which
lead to the Mermin-Wagner-Coleman results \cite{ZJ} for
a continuous symmetry. A criterion to establish the
possibility of spontaneous symmetry breaking is to consider
the ratio
\bqn
r_{c}={Z(\alpha)\over Z(0)} = e^{-\Delta\Gamma}
\eqn
($\Gamma$ is the euclidean action)
between the partition function infinitesimally rotated after an
operation of the symmetry group, and the unrotated one.
To allow for spontaneous symmetry breaking, this ratio has to vanish
in the thermodynamical limit \cite{ZJ}, and thus
$\Delta\Gamma = \Gamma (\alpha)-\Gamma (0)$
has to diverge in the thermodynamical limit.
The opposite (i.e. a finite $\Delta\Gamma$)
would imply that, being the system initially
at one minimum, it would have a non zero probability
to make a transition to another degenerate minimum. This
would ensure the order parameter to be zero \cite{ZJ}.
Actually, only the kinetic term contributes to $\Delta\Gamma$,
which at zero temperature implies, in $D$ dimensions
\bqn
\Delta\Gamma\sim \int (\partial_{\mu} \alpha)^{2} d^{D}x ~
\sim L^{D-2}
\label{dgamma}\eqn
which shows that if $D\leq 2$ no spontaneous symmetry breaking
occurs.
At finite temperature the time dimension is bounded
and the difference in the actions becomes
\bqn
\Delta\Gamma_{\beta}\sim
\int_{0}^{\beta} dx_{0}\int (\partial_{\mu} \alpha)^{2} d^{D-1}x
\sim \beta ~ L^{D-3} ~ \sim ~ \beta ~L^{d-2}
\label{dgammab}\eqn
where $d$ is the number of spatial dimensions. This phenomenon
is much the same as the dimensional reduction at high temperatures. There,
one is considering the high-temperature limit with the spatial
dimensions bounded, here the large spatial-dimension limit
at fixed temperature. Thus a dimensional reduction occurs
whenever the ratio ${x^{i}}/\beta \ra\infty$ $(i = 1, 2,....(D-1))$.\hfb
Finally, if the theory has an internal symmetry group $O(N)$, a $N$
factorizes in the previous expressions
\bqn
\Delta\Gamma \sim N\beta L^{d-2}
\label{dgam}\eqn
and one is lead to discuss in addition the implication of a
$N\rightarrow\infty$
limit. For instance, by allowing
$N\rightarrow\infty$ as $L^{\xi}$, to make $\Delta\Gamma$
divergent for $L\rightarrow\infty$ it would be enough to take
$\xi>2-d$. Thus if $d=1$, the necessary condition is $\xi>1$,
namely that $N$ goes to infinity faster than $L$. \hfb
The large distance behaviour
of the free propagator for massless bosons in $D=d+1$\hfb
dimensions at finite temperature is dominated by
$T\times$(free massless propagator in $d$ dimensions at
zero temperature) (see Appendix D)
\bqn
D_{\beta}({\bf x})\sim T\int {d^{d}k\over (2\pi)^{d}}
{e^{i{\bf k}\cdot {\bf x}}\over {\bf k}^2};~~~~~{\rm for}~
|{\bf x}|\rightarrow\infty
\label{dbeta}\eqn
Thus, as the IR leading term of the action for transverse
fluctuations has the general form (\ref{dgamma}), the expression
(\ref{dbeta}) leads to the same statement of eq. (\ref{dgammab}). In
presence of an internal symmetry group $O(N)$, a
factor $1/N$ appears in (\ref{dbeta}), and then, by putting an IR cutoff
$\sim 1/L$, one finds that in the IR limit the fluctuations
go as
\bqn
D_{\beta}\sim {T\over N~L^{d-2}};~~~~~ {\rm for}~k\rightarrow 0
\eqn
which gives the same information as (\ref{dgam}).\hfb
The considerations about the $N\rightarrow\infty$ limit
are of course not completely satisfactory, as the limit appears
a bit tricky and not physically clear.
However, as the $1/N$ expansion is non-analytic \cite{RW}, to consider the
$N\rightarrow\infty$ limit looks rather as moving to a new distinct
model.\hfb

To summarize, several authors \cite{Ja}\cite{Wo}\cite{KK}\cite{Kl}
have considered the massless Gross-Neveu model
in view of its possible similarity to 3+1 QCD taken in the chiral limit.
Consequently they have neglected the role played by fluctuations,
which can destroy the order at low dimensionality, but should not be so
crucial in 3+1.
We also believe that the study of the model in the mean field approximation
may represent a good guide to the thermodynamics of chiral symmetry
restoration in 3+1 QCD models.
However, both for the discrete and continuous symmetry model, we consider
important to add a non vanishing bare mass term.
This allows to escape
the Mermin-Wagner-Coleman theorem, and besides it
should represent a more realistic simulation of the QCD phase transition.
We have already studied the behaviour of the condensate at
finite temperature and density in a massive QCD-like theory in the mean field
approach \cite{BC}, with results qualitatively equivalent to those
presented here. In this context it is
interesting to study further properties of
the Gross-Neveu model, where the technical complexity is largely reduced.\hfb
We stress again that, differently from QCD, where the zero mass limit
can be safely taken, in the present case the mass has to be kept finite.
In the continuous symmetry model, for instance,
the illness of the zero mass limit is evident by the drastic change in the
long distance behaviour of the free
correlation function, which passes
from an exponential decay to a logarithmic (or linear) divergence.
We remark that a four fermion model in 2+1 dimensions would a priori not be
a good candidate to simulate QCD, as chiral symmetry can be properly
defined only in an even number of space-time dimensions. Thus it
is not surprising that the phase diagram for symmetry restoration
is very different in that case \cite{RW}.\hfb
In the following we will refer
to the mean field results for the massless case only to better clarify
the results of the massive case.

\sect{The massive Gross-Neveu model - Discrete symmetry}

In this section we examine the Gross-Neveu
model when the discrete chiral symmetry is explicitly
broken by adding a fermion mass term to the Lagrangian (\ref{gnlag})
\bqn \LL = {\psib}(i{\hat {\partial}}-M)\psi
+{1\over 2}g^{2}({\psib}\psi)^{2}
\eqn
As already mentioned for the massless case
the model is studied by introducing in the Lagrangian a
$\sigma$ field satisfying the classical
equation of motion $\sigma=g\psib\psi$.
In presence of a mass term it is more
convenient to redefine the $\sigma$ field by shifting it
by a constant $\sigma\ra \sigma+M/g$.
Then the Lagrangian (apart from constant terms) reads
\bqn \LL=\psib i\dslash\psi-\unmez\sigma^2+g\sigma\psib\psi-M{\sigma\over g}
\eqn
Let us first consider the model at vanishing temperature and density.
To study the problem of spontaneous symmetry breaking
(SSB) a suitable tool is
the effective potential $V(\sigma)$,
which has to be minimized with respect to $\sigma$ in order to extract the
physically relevant quantities.
As anticipated in the previous section, we consider the model in the
infinite $N$ limit. Following ref.\cite{DJ}, the effective potential
is obtained as
\bqn
V(\sigma)=\unmez\ss+iN\intkk\ln(\gg\ss-k^{2})+M{\sigma\over g}
\eqn
The renormalization can be carried out by
adding a counterterm  $\delta Z \sigma^{2}/2$
and by imposing the condition
\bqn
d^2 V/d\sigma^2|_{\sigma=\sigma_{0}}=1
\eqn
Then we have
\bqn V(\sigma)
=\unmez\ss+{\lambda\ss\over 4\pi} \bigg[ \ln\bigg({\ss\over\sso}\bigg)
-3\bigg]+M{\sigma\over g}\label{pot1}\eqn
The subtraction parameter $\so$ is arbitrary.
A change in $\so$ is equivalent to a change in $g$, to scaling the
$\sigma$ field, and to rescaling  the bare
mass $M$. Thus the effective potential
must obey the renormalization group equations

\bqn\bigg( \so{\partial\over\partial\so}+\beta (g)
{\partial\over\partial g}-\sigma\gamma_{\sigma}
(g){\partial\over\partial\sigma}
+M\gamma_{M} (g){\partial\over\partial M}\bigg) V(\sigma,\so,g,M)=0
\label{reno}\eqn

If the potential in eq.(\ref{pot1}) is inserted into eq.(\ref{reno})
we find
\bqn \beta(g)&=&-{Ng^{3}/(2\pi)\over 1+Ng^{2}/(2\pi)}\nn\\
\gamma_{\sigma}&=&{\beta(g)\over g}\nn\\
\gamma_{M}&=&2~\gamma_{\sigma}\label{renobeta}\eqn
The proportionality between $\beta(g)$, $\gamma_{\sigma}(g)$
and $\gamma_{M}(g)$ arises from the fact that to order $1/N$
there is no renormalization of the wave function for the $\psi$
field at the one loop level. Therefore the renormalization of
$\sigma$ is determined by the term $g\sigma{\bar\psi}\psi$,
that is from the renormalization of $g$.\hfb
Let us recall that for $M=0$ the
symmetry is spontaneously broken and the
fermion acquires a dynamical mass
\bqn m_{0}=-g\hat{\sigma}=g\so\ee^{1-\pi/\lambda}\label{massao}\eqn
where ${\hat \sigma}$ is the absolute minimum of the effective potential.
Eliminating the dependence of the effective potential on $\so$ by
using the relation (\ref{massao}) between $\so$ and $m_0$,
the expression (\ref{pot1}) can be put in a manifestly invariant form

\bqn V(\sigma)={N \over{4\pi}}\gg\sigma^2 \bigg[
\ln \bigg({g\sigma \over \mo}
\bigg)^2-1\bigg]+g\sigma {M\over g^2 }
\eqn

We now consider the generalization to finite temperature $T$ and
chemical potential $\mu$.
By using standard techniques, one can derive the effective
potential

\bqn V(\sigma,T,\mu)&=&V_{0}(\sigma)+V_{\beta}(\sigma,T,\mu)\nn\\
&=&{N \over{4\pi}}\gg\sigma^2 \bigg[
\ln \bigg({g\sigma \over \mo}
\bigg)^2-1\bigg]+ ~B~ + g\sigma {M\over g^2 }\nn\\
&&-{N\over\beta}\intkl \bigg[\ln ( 1+{\rm e}^{\dis
-\beta({{\sqrt{k^2 + g^2 \sigma ^2}+\mu}})})
+ (\mu\ra  -\mu)\bigg]\label{epot}\eqn
As the pressure equals the negative of
the effective potential evaluated
at the physical points, the equation of state for the system
is given by
\bqn P=-V(T,\mu,{\hat\sigma} (T,\mu,M))\eqn
To normalize to vanishing pressure (and energy density,
in the limit $M\rightarrow 0$) at zero temperature and chemical potential,
we have inserted in the expression for the effective potential (\ref{epot})
the "bag term" $B=Ng^2{\hat\sigma}^2 (0,0,0)/4\pi$
(${\hat\sigma} (0,0,0)={\hat \sigma}=- m_{0}/g)$.
Finally, it is convenient to introduce dimensionless quantities

\bqn
\omega \equiv  {{g\sigma}\over\mo}\quad\quad
\VV \equiv  {\pi\over{N \mo^2}}V\quad\quad
\alpha\equiv {\pi M\over \lambda\mo}\eqn
\bqn
r\equiv T/\mo\quad\quad \eta=\mu/\mo \quad\quad y=k/\mo
\eqn
and
\bqn \VV( \omega)&=&\VV_0 (\omega)+ \VV_{\beta} (\omega,r,\eta)\nn \\
&=& {1\over 4} \omega ^2 ( \ln \omega^2 -1)+{1\over 4}+\alpha\omega\nn\\
&&-r\int_{0}^{\infty} dy \Big[\ln ( 1+{\rm e}^{\dis
-{1\over r}({{\sqrt{y^2 + \omega ^2}+\eta}})})
+ (\eta\ra  -\eta)\Big]\label{adpot}\eqn
which, at the absolute minimum ${\hat{\omega}}$,
gives the dimensionless pressure
\bqn p={\pi\over N m_{0}^2} P=-\VV ({\hat\omega})\label{adpress}\eqn
At zero temperature the effective potential (\ref{adpot}) becomes
\bqn
\VV(\omega)_{r=0}&=&\VV_{0}(\omega)+\VV_{\beta}(\omega,r=0,\eta)\nn\\
&=&{1\over 4} \omega ^2
( \ln \omega^2 -1)+{1\over 4}+\alpha\omega\nn\\
&&+ {1\over 2} \theta (\eta^2 -\omega ^2)[-\eta\sqrt {\eta^2 - \omega ^2}
+\omega^2 \ln ({\sqrt {\eta^2 - \omega ^2} +\eta})-{1\over 2} \omega^2
\ln \omega^2]
 \label{pots}
\eqn

In the following we shall consider the phase diagram and
the equation of state for small values of the symmetry breaking
parameter $\alpha$.

\noindent
\subsection{Phase diagram and thermodynamics}

At $r=\eta=0$ the symmetry is explicitly broken,
and the absolute minimum of the effective potential is on the negative
$\omega$-axis. It appears in correspondence of the point
$\hat{\omega}\sim -(1+\alpha)$, and it becomes deeper and
deeper for increasing $\alpha$, as shown in Fig. 1.

At finite $r$ and $\eta$ the symmetry still remains broken
due to the presence of the breaking term $\alpha\omega$.
Nevertheless, if the mass parameter is small
($\alpha\ll 1$ but different from zero), the system at
finite temperature and density exhibits a line of transitions.
To be more specific, let us start by looking at the behaviour
of the condensate in the plane of chemical potential and temperature
$(\eta,r)$. To describe the results we find convenient to
first recall the phase diagram which derives from the study of the absolute
minima of the effective potential (\ref{adpot}) with $\alpha=0$, already
found in ref. \cite{Wo} (see Fig. 2).
There is a tricritical point $P_{t}=(\eta_{t},r_{t})=(0.608,0.318)$,
separating a line of second order
phase transitions $L_{II}$ starting from the point $(0,r_{c})=(0,0.567)$
and ending at $P_{t}$, from a first order line $L_{I}$ starting from
$P_{t}$ and ending at the point $(\eta_{c},0)=(\sqrt{2}/2,0)$.
In the present case, with $\alpha\not= 0$, by moving
along vertical lines in the $(\eta,r)$ plane, we find the following behaviour:
for chemical potentials higher than a (mass dependent)
value $\eta_{t}(\alpha)$ the condensate undergoes a finite discontinuity
at some (mass dependent) critical temperature and drops
to a value proportional to the bare mass $\alpha$.
Furthermore these critical temperatures decrease for increasing
chemical potentials. Thus the
system still allows for "first order lines" $L_{I}(\alpha)$
starting from points $(\eta_{t}(\alpha),r_{t}(\alpha))$ and ending at points
$(\eta_{c}(\alpha),0)$. On the other hand, for $\eta<\eta_{t}(\alpha)$, the
condensate decreases continuously for growing temperatures and thus
it is no longer possible to extract a line of second order critical points.
Nevertheless there still exists a sharp interval of temperatures in which
the condensate decreases steeply to a small value (proportional to $\alpha$).
This simple result can be summed up by saying
that by taking into account a small bare fermion mass, the phase
diagram of the discrete chiral Gross-Neveu model at finite
temperature and density is a perturbation of the naive mean field
solution of ref. \cite{Wo}, found in the chiral limit.
Although the last one can suffer for the presence of non constant field
configurations which dominate in the thermodynamical limit, the addition
of a bare mass is sufficient to eliminate these complications.
The $1/N$ corrections should not destroy the qualitative picture
in this case. In Fig. 3 we show the comparison between the phase
diagram for $\alpha=0$, for $\alpha=0.01$ and for $\alpha=0.1$.\hfb
In Fig. 4 we show the condensate behaviour vs. temperature
at zero chemical potential for $\alpha=0,\;0.01,\;0.1$.
We see that for
$\alpha\ll 1$ the condensate behaviour is still reminiscent of a second order
phase transition. The same holds for any $\eta<\eta_{t}(\alpha)$.
The low temperature behaviour can be derived analytically
and it is approximately given by~
\bqn \hat{\omega}\simeq -(1+\alpha)
+\sqrt{2 \pi}\bigg(\sqrt{r}-{\alpha\over\sqrt{r}} \bigg)
\ee^{-1/ r}
\eqn\hfb
In Fig. 5 we plot the condensate behaviour vs. chemical potential
at zero temperature for $\alpha=0,\;0.01,\;0.1$, as derived from the effective
potential given by eq.(\ref{pots}).
In this case the critical chemical potential can be evaluated analytically
for $\alpha\ll1$. It turns out to be
\bqn \eta_{c}(\alpha)\sim{\sqrt{2}\over 2}(1+2\alpha)\eqn

Let us now discuss the equation of state. From eqs.(\ref{adpress})
and (\ref{adpot}) we have
\bqn
p(r,\eta)&=&{{\hat\omega}^2\over 4}\left(1-{\rm ln}{\hat\omega}^2\right)
-{1\over 4}~-\alpha {\hat\omega}\nn\\
&&+~r\int_{0}^{\infty}dy~{\rm ln}\Big[1+e^{-(\sqrt{y^2+
{\hat\omega}^2}+\eta)/r}\Big] ~ + ~ (\eta\rightarrow -\eta)\label{press}\eqn
(we recall that the minimum ${\hat\omega}$
depends on $r,~ \eta$ and $\alpha$).\hfb
The fermion number density and the entropy density are obtained by
the derivatives of the pressure $P$ with respect to the chemical potential
and temperature respectively. However it is convenient to define dimensionless
number- and entropy-densities
\bqn n&&={\partial p\over\partial \eta}\nn\\
&&=\int_{0}^{\infty}{dy\over 1+e^{[\sqrt{y^2+{\hat\omega}^2}-\eta]/r}}
{}~ - ~ \left(\eta\rightarrow - \eta\right)\label{enne}\eqn
and
\bqn s&&={\partial p\over\partial r}\nn\\
&&= \int_{0}^{\infty}\left[\left({1\over r}\right)
{\sqrt{y^2+{\hat\omega}^2}+\eta\over
1+e^{[\sqrt{y^2+{\hat\omega}^2}+\eta]/r}}+{\rm ln}
\Big(1+e^{-[\sqrt{y^2+{\hat\omega}^2}+\eta]/r}\Big)\right]+
\Big(\eta \ra\ -\eta\Big)\label{esse}\eqn
which are the true quantities rescaled by a factor $\pi/(Nm_{0})$.
Consistently, the dimensionless energy density is given by the expression
\bqn \varepsilon &&= - p + r {\partial p\over\partial r} +
\eta{\partial p\over\partial \eta} \nn\\
&&= - p + r s + \eta n \nn\\
&&={{\hat\omega}^2\over 4}\left({\rm ln}{\hat\omega}^2-1\right)
+{1\over 4}~+\alpha {\hat\omega}~ + ~ \int_{0}^{\infty}
{\sqrt{y^2+{\hat\omega}^2}\over
1+e^{[\sqrt{y^2+{\hat\omega}^2}+\eta]/r}}
{}~+~\left(\eta\rightarrow - \eta\right)
\label{epsi}\eqn
As we have previously done for the phase diagram, it is useful
first to refer to the results deriving from (\ref{adpot}) with $\alpha=0$.
In this case the exact values of $p,n,s,\varepsilon$
can be derived analytically at $r=\eta=0$ and for $r>r_{c}(\eta)$
(where $r_{c}(\eta)$ is a point of the curves
$L_{II}(\alpha=0)$ or $L_{I}(\alpha=0)$).
In fact, for $\alpha=0$ and $r=\eta=0$, from (\ref{adpot}), one has
${\hat\omega}=1$, and thus from eqs.(\ref{press})-(\ref{epsi})
$p=s=n=\varepsilon=0$, whereas for $r>r_{c}(\eta)$,
one has
${\hat\omega}=0$, and thus the integrals in eqs.(\ref{press})-(\ref{epsi})
can be exactly evaluated, giving
\bqn
p &=& {\pi^2 r^2\over 6} + {\eta^2\over 2} - b \label{pression}\\
s &=& {\pi^2 r\over 3} \label{entropia}\\
n &=&  \eta\label{dens}\\
\varepsilon &=& p+2b \label{energ}
\eqn
(where $b=1/4$ is the dimensionless bag term).\hfb
In Fig. 6 we show the curves $\varepsilon/r^2$ and $p/r^2$ versus
$r$, for $\eta=0$ and $\alpha=0$.
Notice that for $r>r_{c}\simeq 0.57$ the two
curves approach asymptotically
the constant value $K=\pi^2/6$. In fact their expression in this
region is given by
\bqn {p\over r^2} &=& {\pi^2\over 6} + {\eta^2\over 2r^2} -{b\over r^2}\\
{\varepsilon\over r^2} &=&{\pi^2\over 6} + {\eta^2\over 2r^2} +{b\over r^2}
\eqn
and thus for $\eta=0$, the two curves in this
"Stefan-Boltzmann regime", are symmetrical with
respect to the dashed line depicted in Fig. 6.
Notice also that since both $\varepsilon$ and $p$
are exponentially vanishing for $r\ra 0$, their ratio to $r^2$ is
vanishing too.\hfb

In Fig. 7 there are the same quantities as in Fig. 6,
for $\eta=0.63>\eta_{t}(\alpha=0)\simeq 0.608$.
Notice that in this case the curve of the energy
density shows a latent heat as the phase transition is first order, and
that the two curves are no longer symmetrical with respect to the
asymptotic value $K$.\hfb
For $\alpha\not=0$, the pressure and energy density at
$r=\eta=0$ are corrected to $p\simeq\alpha$ and
$\varepsilon\simeq -\alpha$. This is expected
because the energy density has to be proportional
to the condensate, which means
$\varepsilon=\alpha\hat{\omega}\simeq -\alpha$ since
$\hat{\omega}\sim -(1+\alpha)$ at $r=\eta=0$.
A consequence is that at low temperatures the ratios
$p/r^2$ and $\varepsilon/r^2$ are now divergent. Thus, to construct
curves analogous to those in Figs. 6 and 7
(where $\alpha=0$) for the massive case, we have to subtract the divergence.
Actually in Figs. 8 and 9 we show
$[\varepsilon +\alpha]/r^2$ and
$[p-\alpha]/r^2$ vs. $r$ at fixed $\eta$ with
$\eta=0<\eta_{t}(\alpha)$ and $\eta=0.65>\eta_{t}(\alpha)$
respectively, for $\alpha=0.01$. Notice that the latent heat
is slightly  reduced
with respect to the massless case.\hfb

We now come to alternative representations of the phase diagram,
different from those of Figs. 2 and 3.
We consider the planes $(1/n,p)$ and $(n,\varepsilon)$, of
inverse fermion density-pressure and of fermion density-energy density,
which are quantities of more direct physical
interpretation than $r$ and $\eta$.\hfb
Let us start from the isotherms in the $(1/n,p)$ plane
for the massless model, in Fig. 10.
These van der Waals type curves are readily obtained in the following way:
the critical line $L_{II}$ is the mapping
of the corresponding curve in the $(\eta,r)$ plane (see Fig. 2),
whereas the line $L_{I}$ has split into two parts, corresponding to
the value as of the density $n$ at the two degenerate minima at
the critical chemical potential at a given temperature for
$r<r_{t}$. The region inside is the coexistence region.
Notice that it is not necessary to follow
the Maxwell procedure to draw
the isotherms, as all the informations are naturally included in the formalism.
In fact the value of the pressure at the
boundaries of the coexistence region for a given temperature
is directly the negative of the value of the effective potential at the two
degenerate minima at the critical chemical potential.
The "liquid phase" corresponds
to the region where "the symmetry is restored" and thus where
the absolute minimum is ${\hat\omega}=0$, and the "gas phase"
corresponds to the region
where "the symmetry is broken", or ${\hat\omega}\not= 0$.
The intermediate values of $n$ are of course not accessible from the
minima of the effective potential as they only give informations concerning
the pure phases. They are however physically accessible, in analogy,
for instance, with the phenomenon of the melting of ice at zero temperature.
The supercooled and overheated parts of the isotherms can be
easily obtained by following respectively the histories of the minima
at the origin and out of the origin, after they cease to be
absolute minima. The unstable region is
obtained by following the maximum in between, since it appears together
with one of the two minima until it merges with the other.
The picture obtained in this way obeys the Maxwell construction which
is simply derived from the equality of the chemical potentials at the
edge of the two phases.\hfb
The isotherms in the symmetric phase have a simple analytical expression for
$\alpha=0$. In fact, from eqs. (\ref{pression}) and (\ref{dens})
which is $n=\eta$ in the
symmetric phase, one has
\bqn
p={\pi^2 r^2\over 6}+{1\over 2 (1/n)^2}-b ~~~~~~~~~~ (symmetric~phase)
\eqn
The isotherms in the "gas phase" have been obtained numerically.\hfb
The curve $\gamma$ in Fig. 10 is the isotherm at zero temperature,
and therefore it defines
the edge of the accessible region. From the point
$C=(1/n_{c},0)=(\sqrt{2},0)$ it coincides with the horizontal
axis (up to infinity
as $n=0$ for $\eta<\eta_{c}$). Thus the line $\gamma$ has the simple expression
\bqn
p= \theta \left(\sqrt{2}- {1\over n}\right){1\over 2}\left({1\over (1/n^2)}
-{1\over 2}\right)
\eqn
In Fig. 11 are shown the isotherms in the plane of inverse density and
pressure for $\alpha=0.01$, as derived numerically.
The change in the shape of the coexistence region, close to the
critical point $P_{t}$, is well understood by classical
arguments of a Landau expansion (see Appendix A) \cite{DL}.\hfb
As one can see, the description for $\alpha\not=0$ ($\alpha\ll 1$)
is strictly related to that for the massless case. This is not surprising
as the model admits a Landau expansion which does not involve problems
related to long range fluctuations if $\alpha\not=0$.
Thus the system still shows up in two different phases below
the critical pressure $p_{t}$.
The situation for $\alpha\not= 0$ is much the same as for
the vapour-liquid transition for water. It is possible
to obtain Clausius-Clapeyron-like equations from the equality of the pressure
at the edge of the two phases (see ref.\cite{NLW}).\hfb

Finally we come to the phase diagram in the plane
$(n,\varepsilon)$, starting from the case $\alpha=0$ in Fig. 12.
The picture clearly indicates the different regions of chiral symmetry
breaking and restoration, and the region of coexistence, by the mapping of the
corresponding lines in Fig. 10. Here the only isotherm drawn is
the line $\gamma$, which is that for zero temperature, and thus it defines the
edge of the accessible region (as in Fig. 10). It has the simple
expression
\bqn
\varepsilon=\theta \left({\sqrt{2}\over 2}-n\right) {\sqrt{2}\over 2} n +
\theta \left(n-{\sqrt{2}\over 2}\right)\left({1\over 4}+{n^2\over 2}\right)
\label{epsteta}\eqn
The second term in the r.h.s. is obtained from eqs. (\ref{pression}),
(\ref{dens}) and
(\ref{energ}) by taking into account that $n_{c}=\eta_{c}=\sqrt{2}/2$
and $b=1/4$.
As far as the first term is concerned, let us first
recall that, at zero temperature (see also the second line of
eq.(\ref{epsi}))
\bqn\varepsilon=-p +\eta n \eqn
As the phase transition at $r=0$ is first order, the previous formula directly
gives the latent heat in passing from one phase to the other
\bqn \rm{disc}~\varepsilon = \eta_{c}~\rm{disc}~n\eqn
(namely the coordinates of the point $C=(\sqrt{2}/2,1/2)$).
By allowing for the intermediate values of $n$ (which correspond to the
physical values in the mixed phase) and taking into account that
$p_{c}(\eta_{c},r=0)=0$,
whereas the "pure phase" values of $n$ are $n_{1}=0$, $n_{2}=\sqrt{2}/2$,
one gets a straight line
\bqn \varepsilon=-p_{c}+\eta_{c} n= \theta \left({\sqrt{2}\over 2}-n\right)
{\sqrt{2}\over 2}~n \eqn
which is the first term in the r.h.s. of eq. (\ref{epsteta}).\hfb
In Fig. 13 there is the phase diagram in the $(n,\varepsilon)$ plane
for $\alpha=0.01$. Also from this picture (as in Fig. 11),
apart from the
obvious disappearing of the line $L_{II}$, the strong effect of the
addition of a bare mass on the shape of the curve in the vicinity
of the point $P_{t}$ becomes evident.\hfb

\sect{Continuous symmetry}
In this section we consider the
massive Gross-Neveu model described by the Lagrangian
\bqn \LL= {\psib}(i{\hat {\partial}}-M)\psi
+{1\over 2}g^2({\psib}\psi)^2-\unmez g^2(\psib\gamma^{5}\psi)^2
\eqn
which for $M=0$ is invariant under the continuous
chiral transformation (\ref{transfcont}). With\hfb
$M\neq0$ the symmetry is broken from the beginning
and there is no reason for
the appearance of exact Goldstone bosons or exact massless particles in
general.
Actually, as far as the phase diagram and the thermodynamics is concerned,
it is easy to realize that the study of the model in the present case
leads to the same results shown in the previous section.
In fact, after having introduced scalar and pseudoscalar fields and
integrated over the fermion fields, following the same procedure of the
previous section, we obtain an effective potential of the form
\bqn V(\rho^2 ,\sigma,T,\mu)=
V^{(0)}(\rho^2 )+g\sigma{M\over g^2};
\quad\quad\quad(\rho^2 =\sigma^2 +\pi^2)
\eqn
where $V^{(0)}(\rho^2 )$ has the same form of (\ref{epot}) with $M=0$ and
$\sigma\leftrightarrow\rho$
By imposing the minimum conditions
\bqn {dV\over d\sigma}\bigg|_{(\sigma,\pi)=(\bs,\hat{\pi})}
=0={dV\over d\pi}\bigg|_{(\sigma,\pi)=(\bs,\hat{\pi})}
\eqn
it comes out that
\bqn
2\sigma{\partial V^{(0)}
\over\partial \rho^2}\bigg|_{(\sigma,\pi)=(\bs,\hat{\pi})}+{M\over g}=0
\eqn
\bqn
2\pi{\partial V^{(0)}
\over\partial \rho^2}\bigg|_{(\sigma,\pi)=(\bs,\hat{\pi})}=0
\eqn
and thus $\hat{\pi}=0$, and the effective potential reduces
to the one obtained in Section 3.

\subsection{Pion decay constant}
\indent
Our aim now is to analyse the pion mass and decay
constant behaviours in $T$ and $\mu$ in the small mass limit.
First of all we notice that the axial current
$\jc=\psib \gamma_{\mu}\gc\psi$ formally satisfies the PCAC relation
\bqn \DEMU\jc=2{M\over g}\pi\eqn

Moreover, for dimensional reasons, $\pi$ cannot be identical to the
canonical pion field.
Indeed we expect that the physical pion appears in the Lagrangian
with a term
$c_{0}^2 \fp^2$, where $c_{0}$ has the dimension of a mass.
Therefore we impose that the renormalized propagator for $\fp$
has  the canonical form
\bqn D_{\fp}(x)=\bra T\fp(x)\fp(0)\ket=
\int {d^{2}p\over(2\pi)^2}{i\over p^2 -m^2_{\pi}}\ee^{
\dis ip\cdot x}\eqn
Comparing this expression with $D_{\pi}(x)=\bra T\pi(x)\pi(0)\ket$,
we obtain that $c_{0}$ is
the residue of $D_{\pi}(p^2 )$ at the pole corresponding to the pion mass
\footnote{And thus $gc_{o}$ is invariant under the renormalization group.}
\bqn c_{o}^{2}=ResD_{\pi}(p^2 )|_{p^{2}=\mpi}\eqn
The divergence of the axial current then becomes
\bqn \DEMU\jc={2Mc_{o}\over g}\fp\label{uno}\eqn
Defining in the usual way the pion decay constant
\bqn \bra 0|\jc(0)|\fp\ket\equiv i f_{\pi}p_{\mu}\eqn
we have
\bqn \mpi\fpi={2Mc_{o}\over g}\label{emmeffe}\eqn
\bqn \DEMU\jc=\mpi\fpi\fp\label{due}\eqn
In the soft pion limit\footnote{This is the case in which the
$\pi$ propagator can be approximated as $ic_{o}^2 /(p^2 -\mpi)$.}
 (at $T,\mu=0$)
the pion mass can be extracted in the following way \cite{IZ}
  \bqn m_{\pi}^{2}=-ic_{o}^2 D_{\pi}^{-1}(0)
=c_{o}^2 {\partial^2 V\over\partial\pi^{2}}
\bigg|_{min}={\partial^2 V\over\partial\fp^{2}}
\bigg|_{min}\eqn
where $c_{o}^2 $ is the residue of $D_{\pi}(p^2 )$ on the pole.

At finite $T,\mu$, defining $m_{\pi}^{2}(T,\mu)$ as the pole of the
thermal two-point Green function for the $\fp$ field we obtain
  \bqn m_{\pi}^{2}(T,\mu)\equiv -ic^{2}(T,\mu)[D_{\pi}(0)]^{-1}_{\beta}
=c^{2}(T,\mu){\partial^2 V\over\partial\pi^{2}}
\bigg|_{min}={c^{2}(T,\mu)\over c^{2}_{o}}{\partial^2 V\over\partial\fp^{2}}
\bigg|_{min}\eqn
where $c^2 (T,\mu)$ is the residue of $D_{\pi}(p^2 )$ at the pole.
Using the explicit form of the effective potential one has
\bqn m_{\pi}^{2}(T,\mu)=-c^{2}(T,\mu){M\over g\bs}\eqn
Therefore, from eq. (\ref{emmeffe})  we have
\bqn\fpi(T,\mu)=-2{c_{o}\over c^{2}(T,\mu)}\bs\label{effepai}\eqn
\bqn\mpi(T,\mu)\fpi^{2}(T,\mu)=-4{M\over g^2 }{c_{o}^{2}
\over c^{2}(T,\mu)}g\bs\label{emmeffe2}\eqn
Eq.(\ref{emmeffe2})
is just the Adler-Dashen relation at finite temperature and chemical potential.
By comparing it to the $T=\mu=0$ case
\bqn\mpi\fpi^{2}=-4{M\over g^2 }g\bs\label{adlerdashen}\eqn
we see that (\ref{adlerdashen}) is modified  by the term $c^2 _0
/c^2 (T,\mu)$ which depends explicitly on the temperature
and chemical potential
(beside the implicit dependence contained in $g\bs$).

At this point we can obtain
$m_{\pi}(T,\mu)$ and $\fpi(T,\mu)$
by calculating explicitly the factor $c(T,\mu)$ .
If we remember that the inverse
propagator is defined in general as
\bqn D^{-1}(p^{2})=i(1+\delta Z)-\Pi(p^2 )\label{propagator}\eqn
we can find $c(T,\mu)$
by evaluating the self-energy for the $\pi$ field as given
by the diagram shown in Fig. 14.
Carrying on the calculation as done in App.C we finally arrive at
\bqn{1\over c(T,\mu)^{2}}={N\over4\pi \bs^{2}}\bigg[1
+2\bs^{2}{d\over d\bs^{2}}\int{dq \over 2\omega_{q}}\bigg(
{1\over e^{\beta(\omega_{q}+\mu)}+1}+
{1\over e^{\beta(\omega_{q}-\mu)}+1}\bigg)\bigg]\label{cc}\eqn
where we have defined $\omega_{q}=\sqrt{q^2 +g^2 \bs^2}$.

{
We notice that eq. (\ref{cc}) can also be verified,
at least in the chiral limit, by evaluating
 explicitly the axial current matrix element.
 Indeed, for $m_\pi =0$, we obtain
\bqn
ip_{\mu}\fpi=\bra0|\jc|\fp\ket =-2{c_{0}\over c(T,\mu)^{2}}\bs p_{\mu}\eqn}

The behaviours of $f_{\pi}$ and $m_{\pi}$ in $T$ and $\mu$
are shown respectively in Figs. 15, 16
and Figs. 17, 18.
Notice that, if $\alpha=0$ could be taken as physically
meaningful, $f_{\pi}$ could
be used as a physical signal (it is a measurable quantity) of spontaneous
symmetry breaking (for $T<T_{c}$) and of its restoration (for $T>T_{c}$).
Unfortunately this is not the case, because massless bosons in two dimensions
are not allowed. Anyway we notice that for small values of $\alpha$ we can
still distinguish the transition region where $f_{\pi}$ decreases
very rapidly.

As far as the pion mass is concerned we can say that $m_\pi$
grows up quickly for $T>T_{c}$,\hfb
showing that in this case
one is outside the range of validity of the
soft pion hypothesis and the behaviour shown in Fig. 16 keeps only a
qualitative value. Nevertheless we expect the pion mass to be independent of
the symmetry breaking parameter (supposed be small)
once the symmetry is restored, because in such a case
the main parameter would be the scale of the
theory. On the other hand for
$T<T_{c}$, in the case $\alpha=0$, the symmetry would be spontaneously broken
and  the pion would be the
associated Goldstone boson; the mass that it acquires is thus strongly
dependent on the symmetry breaking parameter $\alpha$. Anyway a more
precise analysis, that  we are not interested to carry on in this paper,
could be carried out by considering  the full dependence of
the pion propagator on $p^2$.\hfb
\vfill
\eject
\noindent
\sect{Discussion and Conclusions}\hfb

Study of QCD at finite temperature and density is a formidable problem both
for analytical methods and for numerical simulations. It may then be of
interest to perform similar studies on simpler models to get an introductory
experience in view of the harder QCD problems and to have some hints at the
type of phenomena that may be present. The most easily treatable problems are
in dimensions lower than $4$, particularly in $2$ dimensions, one of space
one of time. Unfortunately the lowest dimensionality brings out peculiar
features in itself, which have been known since some time, and are
particularly expressed by the contents of the so-called
Mermin-Wagner Coleman theorem.

A number of analogies, among them asymptotic freedom, suggest the Gross-Neveu
model in two dimensions as an interesting candidate in order to carry out
orientation studies of what might happen to the physical QCD problem at
finite temperatures and densities. In its original massless form the model
falls into the theoretical problems caused by its low dimensionality, with
direct consequences on symmetry breaking and phase structure, on the validity
of an expansion in the inverse of the number of flavours
($1/N$ expansion), and on the reliability of a mean field approximation.

The simplest massless Gross-Neveu model has a discrete $\gamma_5$ symmetry,
spontaneously broken at zero temperature and zero chemical
potential, without contradiction of the no-go theorems for one-space and
one-time dimensions.
One expects however symmetry restoration as soon as $T\neq0$, for any finite
value of $N$, through a kink-antikink formation mechanism. One can consider
using a mean field approximation only for the strict $N\rightarrow\infty$
limit, where kinks become suppressed and a nominal phase transition takes
place at finite $T$.

When the chemical potential $\mu$ is also non vanishing, the phase diagram for
$N\rightarrow\infty$ shows a structure involving a tricritical point. Our main
interest is the massive case, for which chiral symmetry is never valid and
strictly the no-go theorems do not apply. The expected suppression of
kink-antikink configurations will restore the validity of the large $N$
limit, and one finds that some first order phase transition survives in the
$T-\mu$ plane in spite of the explicit breaking of chiral symmetry, with a
critical line apparently persisting also for large masses.

The phase diagram for the massive Gross-Neveu model has been shown in
Fig. 3, for different values of the mass parameter. Beyond some
(mass-dependent) value of the chemical potential $\mu$ the fermion condensate
shows a discontinuity at a critical value of $T$, which becomes smaller for
higher $\mu$. One has thus lines of first order transitions.

There are no second order transition lines for finite mass values, but the
condensate variation appears as very steep for small masses.

Corresponding to the first order transitions one finds a latent heat visible
in the diagram for the energy density. The phase diagram can, perhaps more
physically, be regarded on the plane (inverse fermion density)-pressure, or
alternatively on the plane (fermion density)-(energy density). The
isotherms in the former plane have the aspect of the popular Van der Waals
curves of water, showing vapour-liquid transition. These Van der Waals curves
are here directly provided by the formalism, without having to apply
Maxwell construction. The end point of the region of phase coexistence is a
critical point. The massive model admits a Landau expansion, as there are
no problems arising from possible long range fluctuations. From the equality
of the pressure at the borders one obtains Clausius-Clapeyron like relations.

A massless Gross-Neveu model exists also with continuous chiral
symmetry. For vanishing $T$ and $\mu$ uncritical use of the mean field
approximation would lead analytical results translating those of the
corresponding discrete symmetry model. One knows however that there cannot be
Goldstones in two dimensions (Coleman theorem). This is reflected by the
appearance of infrared divergent pion correlations at the immediately non
leading order in the $1/N$ expansion. An almost long range
order, corresponding to power law correlation decrease, and reminiscent of the
Kosterlitz-Thouless behaviour of the $X-Y$ spin model, is however expected,
also on simple dimensional arguments. The " almost long range order " does
not survives at any finite $T$, even for large $N$. Only if
$N\rightarrow\infty$ is intended as constructive definition of a distinct new
model one obtains the mean field description. Such constructed model, where
fluctuations have automatically been suppressed, may not be a consistent
approximation to a massless Lagrangian model.

The $N\rightarrow\infty$ properties can be analyzed by a comparison of the
partition functions after and before symmetry rotation, illustrating the kind
of dimensional reduction occurring in a thermal system, and evidencing at the
same time the role played by the infinite $N$. One can also discuss the large
$N$ limit of the fluctuations in the infrared, through study of the
massless thermal propagator.

{}From the point of view of constructing some treatable approximation to
physical QCD, suppressing fluctuations may bring closer to the real
situation. Independently of this consideration our interest has been centered
on massive models, which strictly escape the no-go theorems of two dimensions.
QCD itself has massive quarks, and the general warning when passing to some
massless limit is that this is certainly more dangerous in any comparable
two dimensional model than in QCD itself. Nevertheless we get from our
exercise the conviction that an overall physical content is included in the
$N\rightarrow\infty$ limit. We had previously studied the chiral phase
transition in QCD at finite temperature and density in a mean field composite
model. The analogy of those QCD results to those provided by the
$N\rightarrow\infty$ study of the massive Gross-Neveu model is quite
remarkable and tends to support the impression of a possibly realistic
phase space structure.

The model with continuous chiral symmetry, broken by the mass term, can easily
be led back to the analogous discrete symmetry model to qualitatively
discuss its phase diagram and its thermodynamics at finite $T$ and $\mu$. A
main point of the continuous symmetry study has been
to discuss the properties of
the would-be Goldstone, which however here can never actualize itself as a
Goldstone because of the low dimensionality. Nevertheless one can again build
up for finite mass a kind of Adler-Dashen
relation at finite $T$ and $\mu$, and
derive the behaviours of the " pion decay constant " and of " pion mass ".
One cannot, of course, use the pion-constant in this case to retrace the
symmetry breaking, however we find that traces of the transition remains still
visible in the finite mass model. In conclusion we can only insist on the
overall impression of the usefulness of the Gross-Neveu model, in spite of its
low dimensionality, to illustrate possible behaviours relevant to chiral
phases of QCD at finite $T$ and $\mu$. We have carried out here a detailed
study of such behaviours.
\bigskip
\appendix{

\subsect{Expansion of the effective potential for small $\alpha$}

We show some useful formulas to derive the results of ref.\cite{Wo}
and to approximate our results when the mass parameter $\alpha$ is small.
We remember that, whilst the effective potential we evaluate is a function
which can be safely approximated for small $\alpha$,
to set $\alpha=0$ is a different matter.
In fact for the latter case one should take into account that the
kinetic term of the effective action produces an ill-defined
propagator.
Thus, every result where $\alpha$ is "physically" taken as vanishing, serves
only to recover some known result or to possibly
suggest what would be expected in a 3+1 dimensional model.\hfb
At finite $r$ and $\eta$ the effective potential is given by
\bqn\VV( \omega)&=&{\pi\over 2\lambda}\omega^2 (1+\delta Z)
-r\sum_{n}\intkl
\ln \Big[\Big((2n+1)\pi r-i\eta\Big)^2 +\omega^2 +k^2\Big]+\alpha\omega\nn\\
&\equiv&\VV^{(0)}+\alpha\omega
\label{potmu}\eqn
We notice that at any finite $r$ (and/or $\eta$),
this expression is analytic in the limit
$\omega\ra0$ (this is not true at $r=\eta=0$, due to a logarithmic
divergence in the $n=0$ term).
Thus we can perform a Landau expansion to reproduce the right
structure of the effective potential near to the origin.
We write
\bqn \VV( \omega^2)\simeq{1\over 2}a \omega^2+{1\over 4}b \omega^4+{1\over 6}c
\omega^6+\alpha\omega ~~~~~~~~~~~~~~~~~~~~~~~~~~(\omega\simeq 0)
\label{landaucoeff}\eqn
where the expansion coefficients are given by
\bqn  a\equiv 2{d\VV^{(0)}\over d\ww}\bigg|_{\omega=0}\; ;\quad
b\equiv 2{d^2 \VV^{(0)}
\over d( \ww)^2}\bigg|_{\omega=0}\; ;\quad
c\equiv{d^3 \VV^{(0)}\over d( \ww)^3}\bigg|_{\omega=0}
\eqn
By carrying out explicitly the renormalization, from (\ref{adpot}), we obtain
\bqn
a=1+\intkl \bigg[{\kk \over{[\kk +1]}^{3\over 2}}+{1\over k}\bigg( {1\over
\eps+1}+{1\over \ems+1}-1\bigg) \bigg]\label{eqauno}\eqn
This expression can be integrated, as shown in App.B. We have
\bqn a=\ln \pi r +\re\bigg[2\psi\bigg({i\eta\over r\pi}\bigg)-
\psi\bigg({i\eta\over 2 r\pi}\bigg) \bigg]\label{eqadue}
\eqn
Following this procedure also for $b$ and $c$ would lead to very
complicated expressions. Anyway, by noticing that ultraviolet
divergencies are present only in the $\ww$-term, we can evaluate
 the coefficients $b$ and $c$ directly from eq.(\ref{potmu}).
Carrying out the integration in
$k$, after having derived once, and then deriving again, we find
\bqn b={1\over 8}{1\over \pi^2 r^2}\re\sum_{n=0}^{+\infty}{1\over [n+1/2-{
i\eta/2\pi r}]^3}=
{1\over 8}{1\over \pi^2 r^2}\re \xi\bigg(3,\unmez-{i\eta\over 2\pi r}\bigg)
\label{eqb}\eqn
\bqn
c=
-{3\over 4}\bigg({1\over 2\pi r}\bigg)^4
\re\sum_{n=0}^{+\infty}{1\over [n+1/2-{
i\eta/2\pi r}]^5}=-{3\over 4}\bigg({1\over 2\pi r}\bigg)^4
\re \xi\bigg(5,\unmez-{i\eta\over 2\pi r}\bigg)
\eqn
Using now the relations (\ref{eqadue}) and (\ref{eqb}), and referring to
Fig. 2
, we can easily obtain the following results
\begin{itemize}
\item The equation for the line $AP_{t}D$
(see Treml in ref.\cite{Kl}), solution of $a=0$
\bqn r={1\over\pi}\exp\bigg\{-\re\bigg[2\psi\bigg({i\eta\over r\pi}\bigg)-
\psi\bigg({i\eta\over 2 r\pi}\bigg) \bigg] \bigg\}\eqn
\item The coordinates of the tricritical point $P_{t}$,
determined by the solutions of the equations $a=0,\; b=0$
\footnote{By inserting such results in the expression for $c$ we can verify
that
$c(r_{t},s_{t})>0$, assuring in this way the validity of the previous result.}
\bqn r_t =0.318  \; ,\quad \eta_t =0.608 \;.  \eqn
\item For $\eta=0$ the effective potential becomes $(\epsilon\equiv\omega/r)$
\bqn
\VV(\omega,r,\eta=0)={1\over 4}\omega^2 (\ln \omega^2 -1)
+\alpha\omega -2r^2 \intkl \ln (1+
\ee^{-\sqrt{k^2 +\epsilon^2}})\eqn
Expanding it in $\epsilon^2$ (see ref.\cite{DJ}) we arrive at
\bqn
\VV(\omega)\simeq\unmez(\ln\pi r-\gamma)\ww+{1\over4}{7\over 8\pi^2 }{\xi(3)
\over r^2}\omega^4+\alpha\omega+\dots\label{Landaut}\eqn
where $\gamma=0.577\dots$ is the Euler constant.
In this case it is sufficient to terminate the expansion
at the order $\omega^4$, because
the fact that $b$ is positive assures that the potential is
bounded from below. For $\alpha=0$ the equation $a=0$ gives us
the critical temperature\cite{Ja}
\bqn T_{c}={m_{0}\over\pi}\ee^{\gamma}\simeq0.567m_{0}\eqn
\end{itemize}

\noindent
It is evident that the effective potential (\ref{potmu}),
since it admits the expansion (\ref{landaucoeff})
with IR finite coefficients, leads to the standard mean field
critical exponents for the description of the
system in the vicinity of the line $L_{II}$ (up
to the tricritical point included).

{
We notice that the equations that determine the tricritical point can be
obtained by another method, based on the analysis of the phase diagram
structure, specifically on the fact that the line $P_{t}E$ is the
envelope
of the solutions outside the origin of the stationarity equation for the
potential. In other words (see Fig. 2),
if we consider such solutions, it turns
out that in the region $AP_{t}DO$ there is one and only one
curve for each point (there
is just one minimum\footnote{Due to the symmetry we are just considering
what happens, for examples, for $\omega\leq 0$.}), while in $P_{t}DE$ there are
two solutions for each point (a maximum and a minimum). Therefore
the place where
such points became coincident defines a curve, tangent to the former, that is
just the line $P_{t}E$.

Such a curve can be parametrized by the value $\omega$ that makes the potential
stationary in the interval $[0,1]$ (by the way we notice that $\omega$ is
a continuous function of $T$ and $\mu$). In particular if $F(r,s,\omega)$ is
the equation for the stationary points (outside the origin),
the envelope curve is
given by
\bqn \left \{ \begin{array}{l} F(r,s,\omega)=0\\ {}\\
{\dis {\partial F(r,s,\omega)
\over \partial \omega}=0} \end{array} \right. \eqn
and the tricritical point can be found by imposing $\omega=0$

These equations (evaluated in $\omega=0$) are formally equal to the equations
$a=0,\; b=0$. In fact if we notice that
\bqn
{\partial\VV\over \partial\omega}
=\omega F=2\omega{\partial\VV\over \partial
\ww}\qquad\qquad
{\partial F\over \partial\omega}
=4\omega{\partial^2 \VV\over \partial( \ww)^2}
\eqn
it comes out that the equation ${\partial\VV/\partial\ww}=0$
determines the stationary points outside the origin, while
$\omega\Big[{\partial^2 \VV/\partial( \ww)^2}\Big]=0$ is just the derivative
of the former. Moreover, the second equation must be valid for every value of
the parameter $\omega$ in the range $[0,1]$, and then it reduces to
${\partial^2 \VV/ \partial( \ww)^2}=0$.

These two equations tell us that the envelope curve is the place where the
curvature in the minimum outside the origin changes (the effective potential
becomes like a parabola, and such minimum disappears), and in the $\omega=0$
limit they reduce to $a=0,\; b=0$.}

\subsect{Coefficient $a$ of the Landau expansion}


In this section we show how to carry out the integrations in eq.(\ref{eqauno})
  and
thus obtain eq.(\ref{eqadue}).
The first equation can be transformed according to
 the following formula
\bqn \lefteqn{ \int^{+\infty}_{0}dy\,f(y)\bigg({1\over\ee^{
y-\lambda}+1}+{1\over\ee^{
y+\lambda}+1}\bigg)=}&&\\&&=\int^{+\infty}_{0}dx(f(x+\lambda)
+f(x-\lambda)){1\over\ee^{
x}+1}-\int^{\lambda}_{0}dx (f(x-\lambda)+f(\lambda-x)){1\over\ee^{
x}+1}\nn\\
&&+\int^{\lambda}_{0}dx\,f(x)\label{landau}\nn\eqn
from which we obtain
\bqn a=1+\int^{+\infty}_{0}dy{y\qd\over[y\qd+\alpha]^{3/2}}
-\int_{\lambda}^{+\infty}
dy{1\over y}+P\int^{+\infty}_{0}dx{2x\over x\qd-\lambda\qd}{1\over\ee^{
x}+1}\eqn
The first two integrations give us
\bqn a=\ln\bigg({2\lambda\over \alpha}\bigg)+\int^{+\infty}_{0}dx{2x\over
x\qd-\lambda\qd}{1\over\ee^{ x}+1}\eqn
The remaining integral can be transformed according to
\bqn P\int^{+\infty}_{0}dx{2x\over x\qd-\lambda\qd}{1\over\ee^{
x}+1}&=&P\int^{+\infty}_{0}dx{2x\over x\qd-\lambda\qd}\bigg({1\over\ee^{
x}-1}-{2\over\ee^{ 2x}-1}\bigg)\nn\\
&&\equiv 2I(\lambda)-4I(2\lambda)\eqn
and
\bqn I(\lambda)&=&P\int^{+\infty}_{0}dx{x\over\ee^{
x}-1}{1\over(x+\lambda)(x-\lambda)}\nn\\[6pt]
&=&\lim_{\epsilon\ra 0}\unmez\int^{+\infty}_{0}dx{x\over\ee^{\dis
x}-1}\bigg({1\over(x+\lambda)(x-\lambda+i\epsilon)}+
{1\over(x+\lambda)(x-\lambda-i\epsilon)}\bigg)\eqn
Using now the result \cite{GR}
\bqn\int^{+\infty}_{0}dx{x\over(x\qd+\beta\qd)(\ee^{
x}-1)}=\unmez\bigg[ \,\ln\bigg({\beta\over2\pi}\bigg)-
{\pi\over\beta}-\psi\bigg({\beta\over2\pi}\bigg)\bigg]
\label{grad}\quad (\re \beta>0)\eqn
and carrying out the limit, we obtain
\bqn
2I(\lambda)-4I(2\lambda)=\ln{\pi\over\lambda}+
\re\bigg[2\psi\bigg({i\lambda\over\pi}\bigg)
-\psi\bigg({i\lambda\over2\pi}\bigg)\bigg]\eqn
from which is easy to get  the wanted result.

\subsect{Pion self-energy}

{}From Fig. 14 we have
\bqn \Pi(p^{2})&=&N(-g)^2 \int_{k}(-1)\tr\bigg[\gc{i\over \hat{p}+\hat{k}-m}\gc
{i\over \hat{k}-m}\bigg]\nn\\
&=&2\lambda\int_{k} {m^2 -k(k+p)\over
\big((p+k)^2 -m^{2}\big)(k^2 -m^2 )}\eqn
By using the usual Feynman parameterization we can write
\bqn \Pi(p^{2})=2\lambda\int_{0}^{1}d\alpha\int_{q}{m^{2}-q^{2}+p^{2}\alpha
(1-\alpha)\over[q^{2}+p^{2}\alpha(1-\alpha)-m^2 ]^{2}}\eqn
In the soft pion limit we can write
\bqn \Pi(p^{2})=\Pi_{o}+p^{2}\Pi_{1}+\dots\eqn
where
\bqn \Pi_{o}=-2\lambda\int_{q}{1\over q^{2}-m^2 }\quad\quad
\Pi_{1}=\int_{q}{1\over (q^{2}-m^2 )^{2}}\eqn
The inverse of the pion propagator residue  on the pole is just $-i\Pi_1 $.
This can be evaluated using the Poisson summation
\bqn
{1\over\beta}\sum_{n} f\bigg[(2n+1){\pi\over\beta}+i\mu\bigg]=
\sum_{n}(-1)^n \int{dq_{o}\over2\pi} f(q_{o}){\ee^{in\beta (q_{o}-i\mu)}}
\eqn
We have
\bqn \Pi_{1}&=&i\lambda\int{dq\over 2\pi}\sum_{n}(-1)^n \int{dq_{o}\over2\pi}
{\ee^{in\beta (q_{o}-i\mu)}\over (q_{o}^{2}+q^{2}+m^{2})^{2}}\nn\\[6pt]
&=&{i\lambda\over4\pi m^{2}}+
2i\lambda\int{dq\over 2\pi}\sum_{n=1}^{+\infty}(-1)^n
 (-1){d\over dm^{2}}\int{dq_{o}\over2\pi}
{\ee^{in\beta (q_{o}-i\mu)}\over (q_{o}^{2}+q^{2}+m^{2})}\eqn
The integral in $dq_{0}$ can be performed using the Cauchy theorem, and the
sum turns out to be just a geometric series. At the end we arrive at
\bqn \Pi_{1}={i\over c^2 (T,\mu)}={iN\over4\pi \bs^{2}}
+2iN{d\over d\sigma^{2}}\int{dq\over2\pi}{1\over 2\omega_{q}}\bigg(
{1\over e^{\beta(\omega_{q}+\mu)}+1}+
{1\over e^{\beta(\omega_{q}-\mu)}+1}
\bigg)\label{selfuno}\eqn

\subsect{Bosonic free propagators}
\noindent
{\bf Zero temperature}\hfb

\noindent
We recall some known results at zero temperature.\hfb
In $D$ euclidean dimensions, the free two-point
function in coordinates space
\bqn
D_{0}(x,m^2)= \int {d^{D}k\over (2\pi)^{D}} ~
{e^{-i k\cdot x}\over
k^2+m^2}
\label{dtransf}\eqn
satisfies the wave equation
\bqn
(\dalamb_{D}-m^2)D_{0}(x,m^2)=- \int {d^{D}k\over (2\pi)^{D}} ~
e^{-ik\cdot x}
=-\delta^{D}(x)
\label{dprop}\eqn
where
\bqn
\dalamb_{D} f = \sum_{\mu=1}^{D}
{\partial\over\partial x^{\mu}}{\partial\over\partial x^{\mu}} f
\eqn
Lorentz invariance requires that $D_{0}(x,m^2)$ depends only  on
$r=\sqrt{x_{\mu}x_{\mu}}$; therefore
(\ref{dprop}) becomes
\bqn
\Big[{d^2\over dr^2}+{D-1\over r}{d\over dr} -  m^2\Big]~D_{0} (r,m^2)=
-\delta^{D}(x)
\eqn
By using the identity
\bqn
{1\over k^{2}+m^{2}}=\int_{0}^{\infty}d\alpha ~ e^{-\alpha (k^2+m^2)}
\eqn
in eq.(\ref{dtransf}), one gets
\bqn
D_{0}(r,m)={1\over (2\pi)^{D/2}} ~ {K_{{D\over 2}-1}(m r)\over
\left[ r^{(D-2)/2}~m^{(2-D)/2}\right]}
\eqn
where use has been made of the representation for the modified
Bessel functions
\bqn
K_{\nu}(z)&=&{1\over 2}\left({z\over 2}\right)^{\nu}
\int_{0}^{\infty} t^{-\nu-1}~e^{-t-z^2/4t} ~ dt\label{bessel}\\
&& |arg ~z|<\pi/2; ~ ~ ~  Re~z^2>0\nn
\eqn
which behave \cite{AS}, for fixed $\nu$ and $z\rightarrow 0$
\bqn
K_{0}(z) ~ &\sim& ~ -~{\rm ln}\left({z\over 2}\right)-\gamma\nn\\
K_{\nu}(z) ~ &\sim& ~ {1\over 2}~\Gamma (\nu)~\left({z\over 2}\right)
^{-\nu} ~ ~ ~ ~ ; ~ ~ ~ ~ ({\rm Re}~\nu~>~0)
\label{bess}\eqn
where $\gamma=0.577...$ is the Euler's constant.\hfb
These formulas allow to recover the well known behaviours
of the free massless propagators in D=2+1 and D= 3+1 at zero temperature
\bqn
D_{0}(r)&=&{1\over 4\pi}{1\over r}\nn\\
D_{0}(r)&=&{1\over 4\pi^2}{1\over r^2}
\eqn
For $D=1+1$, performing the $m\rightarrow 0$
limit with fixed $r$, from (\ref{bess}) one finds that the expression for the
free massless propagator in two (infinite) dimensions
\bqn
D_{0}(r)= -{1\over 2\pi} ~ {\rm ln} (r) -{1\over 2\pi}~{\rm ln}
\left({m\over 2}\right)-{\gamma\over 2\pi}
\label{dzero}\eqn
is logarithmically divergent for large distances.\hfb

\noindent
{\bf Finite temperature free propagators in D=d+1 dimensions}\hfb

\noindent
By working in the imaginary times formalism, the passage from Minkowski
metrics to Euclidean metrics comes out naturally. In fact one has
\bqn
D_{\beta}(x)=\int_{k}~{i\over k^2-m^2+i\varepsilon}~e^{-ik\cdot x}
\eqn
with
\bqn
k^{\mu}&=&(k_{0},{\bf k})=(2\pi i n T,{\bf k}) \nn\\
x^{\mu}&=&(x_{0},{\bf x})=(-i\tau,{\bf x})
\eqn
$-\beta<\tau<\beta$, and
\bqn
\int_{k}={i\over\beta}\sum_{n=-\infty}^{+\infty}
\int_{-\infty}^{+\infty}{d^{d}k\over (2\pi)^{d}}
\label{kint}\eqn
thus ($\beta=1/T$)
\bqn
D_{\beta}(\tau,{\bf x})=~T~\sum_{n=-\infty}^{+\infty}
\int_{-\infty}^{+\infty}{d^{d}k
\over (2\pi)^{d}}{e^{-i 2\pi n T \tau +i {\bf k}\cdot {\bf x}}
\over 4\pi^2 n^2 T^2 +{\bf k}^2+m^2}
\label{dbeta1}\eqn
One wants to evaluate the ${\bf x}$ behaviour in the
massless limit for $D=1+1$, $D=2+1$ and $D=3+1$, and in particular
the large distances behaviour in the massless limit.\hfb
To this end it is sufficient to separate out  the $n=0$ term
in eq.(\ref{kint}) after having set $m=0$\hfb
\bqn
D_{\beta}(\tau,{\bf x})=T\int {d^{d}k\over (2\pi)^{d}}
{e^{i{\bf k}\cdot {\bf x}}\over {\bf k}^2} ~ + ~ {1\over 2\pi^2 T}
\int {d^{d}k\over (2\pi)^{d}}~
e^{i{\bf k}\cdot {\bf x}}~ \sum_{n=1}^{+\infty}
{cos (ny)\over n^2+\alpha^2}
\label{dbetaz}\eqn
where $\alpha\equiv k/(2\pi T)$ and $y\equiv 2\pi T \tau$.\hfb
The sum of the series is
\bqn
\sum_{n=1}^{+\infty}
{cos (ny)\over n^2+\alpha^2}=
{\pi\over 2\alpha}{ch [\alpha (\pi - y)]\over sh (\alpha \pi)}
-{1\over 2\alpha^2}
\label{summ}\eqn
The large distances behaviours are determined by the IR leading term
of (\ref{dbeta1}). It is easy to see, by expanding (\ref{dbetaz}) for small
$k$,
that the contribution of the sum (\ref{dbetaz}) is subleading with
respect to the first term on the r.h.s. in (\ref{dbeta1}).
Thus the free massless propagator in D=d+1
dimensions at finite temperature for large distances behaves simply
as\break
$T\times$(free massless propagator in d dimensions
at zero temperature).\hfb
Let us now evaluate the full ${\bf x}$ dependence for $d=1,2,3$.\hfb

\noindent
{\bf D=1+1}\hfb

\noindent
In this case
\bqn
D_{\beta}(\tau,x)=T\sum_{n=-\infty}^{+\infty}
\int_{-\infty}^{+\infty}{dk
\over 2\pi}{e^{-i 2\pi n T \tau +i k x}
\over 4\pi^2 n^2 T^2 +k^2+m^2}
\label{ddbeta}\eqn
and it is evident that the zero mass limit is badly divergent.
Anyway this expression has a well defined behaviour in $x$ as it can be
easily seen by the fact that its derivative with respect to $x$
is a convergent integral even for $m=0$. \hfb
By using the result (\ref{dbetaz})
the massless propagator can be put in the form
\bqn
D_{\beta}(\tau,x)={1\over 2\pi}\int_{0}^{+\infty} {dk\over k} ~
cos (kx) ~ {ch\Big[k\Big({\beta\over 2}-\tau\Big)\Big]\over
sh\Big(k{\beta\over 2}\Big)}
\eqn
As expected the IR behaviour of this integral is the same as
that of the $n=0$ term of (\ref{summ}).\hfb
Let us now evaluate the $x$ derivative of (\ref{ddbeta})
\bqn
{\partial D_{\beta}(\tau,x)\over\partial x}&=&
{}~ - ~ \int_{0}^{+\infty} {dk\over 2\pi}{sin(kx)
{}~ ch\Big[k\Big({\beta\over 2}-\tau\Big)\Big]\over
sh\Big(k{\beta\over 2}\Big)}\nn\\
&=& -{T\over 2} ~ {sh(2\pi x T)\over ch(2\pi x T)~ - ~cos(2\pi\tau T)}
\eqn
Integrating again one gets
\bqn
D_{\beta}(\tau,x)=C(T) - {1\over 4\pi}~{\rm ln}
\Big[ ch(2\pi T x)-cos(2\pi T \tau)
\Big]
\eqn
where $C(T)$ is an integration constant which does not depend on $x$.
It is easy to verify that in the $T\rightarrow 0$ limit it satisfies
(\ref{dzero}) in the coordinate dependence in two dimensions at
zero temperature, whereas at finite $T$,
for $|x|\rightarrow\infty$ the leading
behaviour is that expected for one dimension
\bqn
D_{\beta}(\tau,x) \sim  -{T |x|\over 2}
\eqn
\hfb

\noindent
{\bf D=2+1}\hfb

\noindent
We must evaluate
\bqn
D_{\beta}(\tau,{\bf x})&=&T\sum_{n=-\infty}^{+\infty}
\int_{-\infty}^{+\infty}{d^2
k\over (2\pi)^2}{e^{-i 2\pi n T \tau +i {\bf k\cdot x}}
\over 4\pi^2 n^2 T^2 + {\bf k^2}+m^2}\nn\\
&=&{T\over 2\pi}\sum_{n=-\infty}^{+\infty}\int_{0}^{+\infty}
{k~J_{0}(kx)~e^{-i 2\pi n T \tau }\over 4\pi^2 n^2 T^2 + k^2+m^2} dk
\eqn
Separating out the $n=0$ term (from here on remember
that $x\equiv |{\bf x}|$), one gets

\bqn
D_{\beta}(\tau,x)
={T\over 2\pi} K_{0}(mx)+{T\over\pi} \sum_{n=1}^{\infty} cos(2\pi n T\tau)
K_{0}(x\sqrt{4\pi^2 n^2 T^2+m^2})
\eqn
The sum of the series for $m=0$ is not a closed form
\bqn
\sum_{n=1}^{\infty} cos(ny)K_{0}(nz)&=&
{\pi\over 2}{1\over \sqrt{z^2+y^2}}+ {1\over 2}\Big[\gamma-{\rm ln}
\Big({4\pi\over
z}\Big)\Big]\nn\\
&&+{\pi\over 2}\sum_{n=1}^{\infty}\Big[{1\over\sqrt{z^2+(2\pi n -
y)^2}}-{1\over
2\pi n}\Big]\nn\\
&&+{\pi\over 2}\sum_{n=1}^{\infty}\Big[{1\over\sqrt{z^2+(2\pi n +y)^2}}-{1\over
2\pi n}\Big]~~~;~~~z>0, y\epsilon R
\eqn
($\gamma=0.577..$ is again the Euler's constant)\hfb
Identifying $y\equiv 2\pi T\tau$, $z\equiv 2\pi T x$, and taking
into account that, as already seen in (\ref{bessel}), for $m\rightarrow 0$
\bqn
K_{0}(mz)\sim -{\rm ln}\left({mz\over 2}\right)-\gamma
\eqn
it follows that for $m\rightarrow 0$ the propagator is given by the
expression
\bqn
D_{\beta}(\tau,x)&\sim & -{T\over 2\pi}{\rm ln}(m)+{T\over 2\pi}
{\rm ln}(T) \nn\\
&&+{1\over 4\pi}{1\over\sqrt{x^2+\tau^2}}\nn\\
&&+{T\over 4\pi}\sum_{n=1}^{\infty}\Big[{1\over\sqrt{T^2 x^2 + (n-T\tau)^2}}
-{1\over n}\Big]\nn\\
&&+{T\over 4\pi}\sum_{n=1}^{\infty}\Big[{1\over\sqrt{T^2 x^2 + (n+T\tau)^2}}
-{1\over n}\Big]
\label{ddbetaz}\eqn
In the second line we see the term which survives for $T\rightarrow 0$, which,
as expected, corresponds to the first term of (\ref{bess}).\hfb
For $T\not= 0$, the dominant term for $x\rightarrow\infty$ comes
from summation of the series. We can use the summation formula \cite{Ol}
\bqn
\sum_{n=a}^{N} f(n)&=&\int_{a}^{N}f(x)dx +{1\over 2}f(a)+{1\over 2}f(N)\nn\\
&&+\sum_{s=1}^{q-1}{B_{2s}\over (2s)!}\Big[
f^{(2s-1)}(N)-f^{(2s-1)}(a)\Big]+R_{q}(N)
\eqn
where $a$,$q$,$N$, are arbitrary integers with $a<N$ and $q>0$, and
\bqn
R_{q}(N)={B_{2q}\over (2q)!}\Big[f^{(2q-1)}(N)-f^{(2q-1)}(a)\Big]
-\int_{a}^{N}{B_{2q}(x-[x])\over (2q)!}f^{(2q)}(x)dx
\eqn
This formula holds if $f^{(2q)}(x)$ is absolutely
integrable over $(a,N)$. The first Bernoulli coefficients are
\bqn
{B_{2}\over 2!}={1\over 12}; ~~~~~ {B_{4}\over 4!}=-{1\over 720}
\eqn
It is easy to see that for $x\rightarrow\infty$ the terms in the first
line of (\ref{ddbetaz}) are dominant and that ${\rm ln}(m)$ is the
only infinite constant for $m\rightarrow 0$. The final result,
giving the large $x$ dependence in the massless limit, can be put in the form
\bqn
D_{\beta}(\tau,x)&=& -{T\over 2\pi}{\rm ln}(m)+g(\tau,x,T)\nn\\
&&-{T\over 4\pi}{\rm ln}\Big[\beta+\tau+\sqrt{x^2+(\beta+\tau)^2}\Big]\nn\\
&&-{T\over 4\pi}{\rm ln}\Big[\beta-\tau+\sqrt{x^2+(\beta-\tau)^2}\Big]
\eqn
where
\bqn
g(\tau,x,T)&\rightarrow& 0 ~~  per ~x\rightarrow\infty\nn\\
g(\tau,x,T)&\rightarrow& {1\over4\pi}{1\over
\sqrt{x^2 + \tau^2}} ~~  for ~T\rightarrow 0
\eqn
and thus, for $x\rightarrow\infty$
\bqn
D_{\beta}(\tau,x)\sim -{T\over 2\pi} {\rm ln}(x)
\eqn
which is, as expected, the behaviour typical of two dimensions.\hfb

\noindent
{\bf D=3+1}\hfb

\noindent
In D=3+1 the massless propagator is well defined even at finite
temperature
\bqn
D_{\beta}(\tau,{\bf x})&=&{1\over \beta}\sum_{n=-\infty}^{+\infty}
\int_{-\infty}^{+\infty}{d^3 k\over
(2\pi)^3}{e^{-i 2\pi n T \tau +i {\bf k\cdot x}}
\over 4\pi^2 n^2 T^2 + {\bf k^2}}\nn\\
&=&{T\over 2\pi^2 x}\sum_{n=-\infty}^{+\infty}
\int_{0}^{\infty}{k~sin(kx)\over 4\pi^2 T^2 n^2 +k^2}~e^{-i2\pi nT\tau}dk
\eqn
By separating the $n=0$ term, and using again the sum of the series in
(\ref{dbetaz}),
one easily arrives to the expression
\bqn
D_{\beta}(\tau,x)
={T\over 4\pi x}~{sh(2\pi T x)\over ch(2\pi T x)-cos(2\pi T \tau)}
\eqn
For $T\rightarrow 0$ it is easy to recover the
four dimensional result already given in the second of (\ref{bess}),
whereas for $T\not= 0$
and $x\rightarrow\infty$ the leading behaviour is that of three
dimensions
\bqn
D_{\beta}(\tau,x)\sim {T\over 4\pi x}
\eqn}

\newpage

\centerline{\bf FIGURE CAPTIONS}
\bigskip
\parbox[t]{2cm}{\bf  Fig. 1:}\parbox[t]{14cm}{Effective potential
                     at $T=\mu=0$ for various values of $\alpha$.}\\

\parbox[t]{2cm}{\bf  Fig. 2:}\parbox[t]{14cm}{Phase diagram for the massless
                             Gross-Neveu model.}\\

\parbox[t]{2cm}{\bf  Fig. 3:}\parbox[t]{14cm}{Phase diagram for the massive
                   Gross-Neveu model.}\\

\parbox[t]{2cm}{\bf  Fig. 4:}\parbox[t]{14cm}{Condensate behaviour in $r$
                              at $\eta=0$ for
                     $\alpha=0$, $\alpha=0.01$ and $\alpha=0.1$.}\\

\parbox[t]{2cm}{\bf  Fig. 5:}\parbox[t]{14cm}{Condensate behaviour in $\eta$
                    at $r=0$ for $\alpha=0$, $\alpha=0.01$
                    and $\alpha=0.1$.}\\

\parbox[t]{2cm}{\bf  Fig. 6:}\parbox[t]{14cm}{Plot of $\varepsilon/r^2$ and
                    $p/r^2$ vs $r$ for $\eta=0$ at $\alpha=0$.}\\

\parbox[t]{2cm}{\bf  Fig. 7:}\parbox[t]{14cm}{Plot of $\varepsilon/r^2$ and
                    $p/r^2$ vs $r$ for $\eta=0.63>\eta_{t}$ at $\alpha=0$.}\\

\parbox[t]{2cm}{\bf  Fig. 8:}\parbox[t]{14cm}
                    {Plot of $[\varepsilon+\alpha]/r^2$ and
                    $[p+\alpha]/r^2$ vs $r$ for
                    $\eta=0$ at $\alpha=0.01$. Remember that
                    $\varepsilon(r=0,\eta=0)\simeq-\alpha$ for $\alpha\ll
1$.}\\

\parbox[t]{2cm}{\bf  Fig. 9:}\parbox[t]{14cm}
                    {Plot of $[\varepsilon+\alpha]/r^2$ and
                    $[p-\alpha]/r^2$ vs $r$ for
                    $\eta=0.65>\eta_{t}(\alpha)$ at $\alpha=0.01$.Remember that
                    $\varepsilon(r=0,\eta=0)\simeq-\alpha$
                    for $\alpha\ll 1$.}\\

\parbox[t]{2cm}{\bf  Fig. 10:}\parbox[t]{14cm}{Isotherms of the massless
                     Gross-Neveu model in the $(1/n,p)$ plane.}\\

\parbox[t]{2cm}{\bf  Fig. 11:}\parbox[t]{14cm}{Isotherms of the massive
                    Gross-Neveu model in the $(1/n,p)$ plane
                     for $\alpha=0.01$.}\\

\parbox[t]{2cm}{\bf  Fig. 12:}\parbox[t]{14cm}{Phase diagram for the massless
                    Gross-Neveu model in the $(n,\varepsilon)$ plane.}\\

\parbox[t]{2cm}{\bf  Fig. 13:}\parbox[t]{14cm}{Phase diagram for the massive
                       Gross-Neveu model in the $(n,\varepsilon)$ plane for\hfb
                       $\alpha=0.01$.}\\

\parbox[t]{2cm}{\bf  Fig. 14:}\parbox[t]{14cm}{Pion self-energy at one loop
                              order.}\\

\parbox[t]{2cm}{\bf  Fig. 15:}\parbox[t]{14cm}{Pion decay constant
                    vs. temperature for $\alpha=0$,
                    $\alpha=0.01$ and $\alpha=0.1$.}\\

\parbox[t]{2cm}{\bf  Fig. 16:}\parbox[t]{14cm}{Pion mass vs. temperature
                    for $\alpha=0.01$ and $\alpha=0.1$.}\\

\parbox[t]{2cm}{\bf  Fig. 17:}\parbox[t]{14cm}{Pion decay constant vs.
                         chemical potential for $\alpha=0.01$.}\\

\parbox[t]{2cm}{\bf  Fig. 18:}\parbox[t]{14cm}{Pion mass vs. chemical
                         potential for $\alpha=0.01$.}

\end{document}